\documentclass{IEEEcsmag}

\usepackage[colorlinks,urlcolor=blue,linkcolor=blue,citecolor=blue]{hyperref}

\usepackage{upmath}

\usepackage{csquotes}
\usepackage{setspace}

\usepackage[many]{tcolorbox}  
\definecolor{main}{HTML}{5989cf}    % setting main color to be used
\definecolor{sub}{HTML}{eeeeee}     % setting sub color to be used
\newtcolorbox{boxC}{
    left=3pt, 
    right=3pt,
    breakable,
      sidebyside align=top,
  sidebyside gap=5pt,
    colback = sub, % background color
    boxrule = 0pt  % no borders
}
\usepackage[absolute,showboxes]{textpos}

\TPMargin{4pt}

\newcommand{\copyrightstatement}{
    \begin{textblock}{0.71}(0.2,1.00)    % tweak here: {box width}(leftposition, rightposition)
         \noindent
         \centering \scriptsize
         Copyright~\copyright~2023 IEEE. Personal use of this material is permitted. Permission from IEEE must be obtained for all other uses, in any current or future media, including reprinting/republishing this material for advertising or promotional purposes, creating new collective works, for resale or redistribution to servers or lists, or reuse of any copyrighted component of this work in other works by sending a request to pubs-permissions@ieee.org.
    \end{textblock}
}
\jvol{XX}
\jnum{XX}
\paper{8}
\begin{document}

\copyrightstatement

\title{Resist the Hype! Practical Recommendations to Cope With Résumé-Driven Development}

\author{Jonas Fritzsch, Marvin Wyrich, Justus Bogner, Stefan Wagner}
\affil{University of Stuttgart, Germany}

\begin{abstract}
Technology trends play an important role in the hiring process for software and IT professionals. In a recent study of 591 software professionals in both hiring (130) and technical (558) roles, we found empirical support for a tendency to overemphasize technology trends in résumés and the application process. 60\% of the hiring professionals agreed that such trends would influence their job advertisements. Among the software professionals, 82\% believed that using trending technologies in their daily work would make them more attractive for potential future employers. This phenomenon has previously been reported anecdotally and somewhat humorously under the label Résumé-Driven Development (RDD). Our article seeks to initiate a more serious debate about the consequences of RDD on software development practice. We explain how the phenomenon may constitute a harmful self-sustaining dynamic, and provide practical recommendations for both the hiring and applicant perspectives to change the current situation for the better.
\end{abstract}

% \begin{IEEEkeywords}
% %
% software development, technology, career, hiring, development, survey, theory
% %
% \end{IEEEkeywords}

\maketitle

\chapterinitial{The future of software} development would have to cope with serious challenges if we adhered to the satirical manifesto of \textit{Résumé-Driven Development} (RDD) \cite{rdd.io2020}: 

\begin{boxC}
\large
\setstretch{1.15}
\noindent
\enquote{\textit{Specific technologies over working solutions, hiring buzzwords over proven track records, creative job titles over technical experience, and reacting to trends over more pragmatic options.}}  
% \normalsize
%-- satirical manifesto about Résumé-Driven Development \cite{rdd.io2020}
\end{boxC}

\noindent
While this is obviously a humorous play on the Agile Manifesto,\footnote{\url{https://agilemanifesto.org}} it entails at least a grain of truth. 
In times of social networks, communities, job portals, and especially career platforms, a software developer's portrait reflecting the professional résumé has become more of a figurehead than ever.
An up-to-date profile on LinkedIn\footnote{\url{https://www.linkedin.com}} comprising the professional career, degrees, obtained certificates as well as knowledge and skills confirmed by colleagues (\enquote{endorsements}) is nowadays rather the rule than the exception.
Thanks to sophisticated search capabilities, those who present themselves appropriately and comprehensively on such platforms will be found first — a %real
treasure trove for headhunters and companies. % as many of us can certainly confirm.

What at first glance looks like an extremely useful tool for applicants and hiring professionals has its downsides as well.
Apart from general concerns about digital abstinence or data protection, the great importance of profiles and résumés being constantly available to %almost 
everyone also leads to an increasing urge to perfect one's own appearance. 
A recent survey among 65,000 software developers~\cite{Stackoverflow2020} showed that knowledge and skills regarding various technologies play an important role in their application process.
While hiring professionals spend on average just 7 seconds looking at an applicant's résumé \cite{Ladders2018}, it may be tempting to try to convince the recruiters' critical eyes through breadth and trendiness of used technologies, often referred to as \enquote{buzzwords} by more cynical contemporaries. 

\section{The Phenomenon of Résumé-Driven Development}
\medskip

\noindent
It appears that \emph{one} possible strategy for building an impressive personal profile in the current field of activity is to work with a wide selection of recent and popular technologies.
Moreover, future employers and jobs may specifically be chosen according to whether they contribute to that goal.

 \begin{figure}[ht!]
    \raggedleft
    \includegraphics[width=0.485\textwidth]{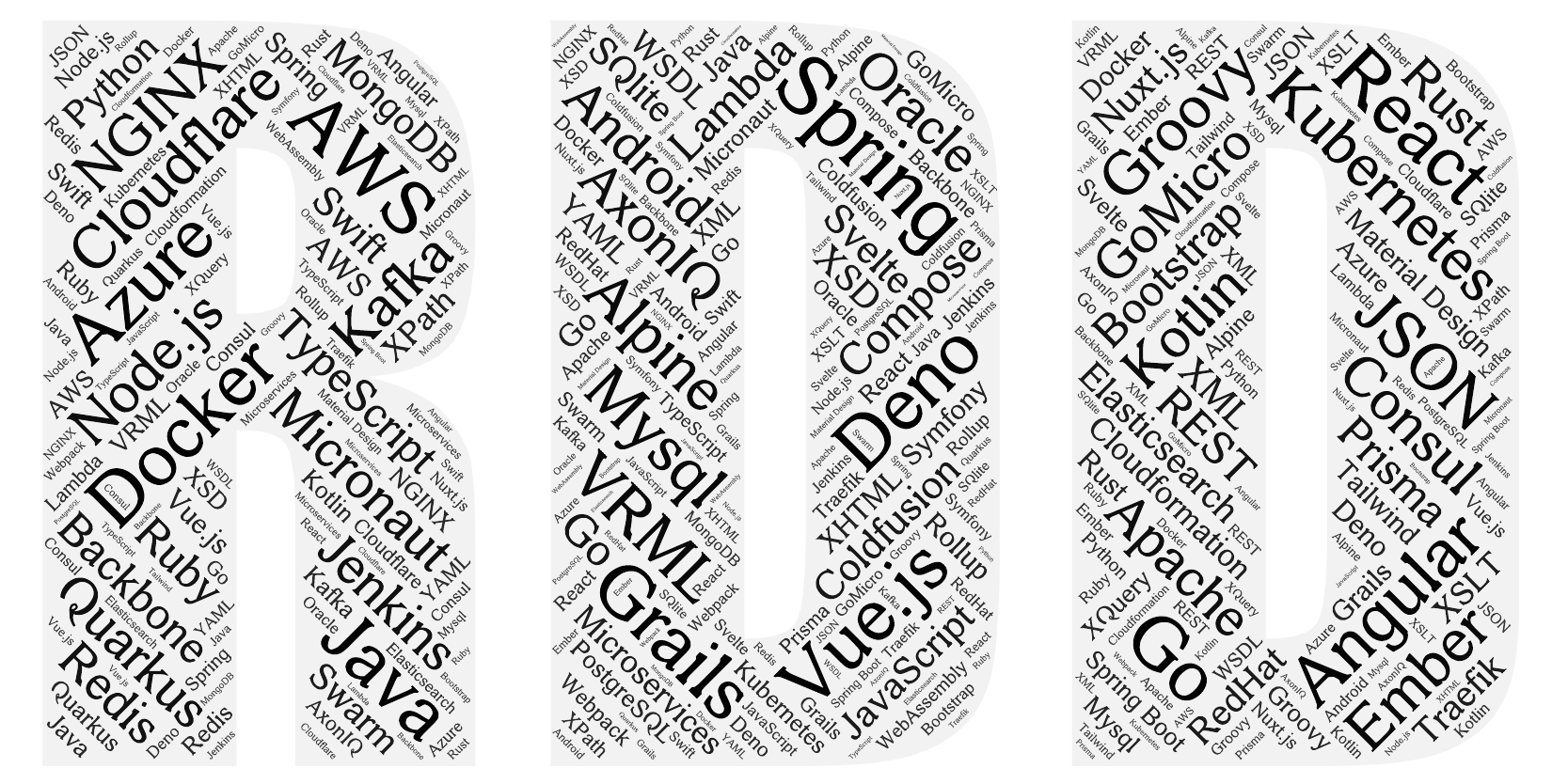}
    \caption{Today's technology landscape is growing faster than ever, bringing forth numerous trends too}
    \label{fig:rdd-word-art}
\end{figure}

\noindent
The essence of a phenomenon termed \textit{Résumé-Driven Development} lies in a focus on current technology trends (some of which we summarized visually in Fig.~\ref{fig:rdd-word-art}) that fill supposed gaps in the applicant's profile, thereby extending it and making it appear more impressive. Focusing on CV attractiveness then inevitably supersedes or replaces project-specific requirements, which should actually be the primary driver to select technologies.

Following RDD, a frontend developer would prioritize, e.g., Angular or React in the current project, if they lack practice with one of these frameworks. If there is too little with microservices in a developer's CV, they will perhaps implement this simple web app using Spring Boot and deploy it as 15 containers in a Kubernetes cluster. One's preference is based on current trends or hypes that look exciting in the résumé. First, however, they may not have been understood in depth and secondly, they often disappear from the market after a short time. 
Here, the regularly appearing Gartner Hype Cycle~\cite{Gartner2021} complemented by ThoughtWorks' quarterly technology radar ~\cite{Thoughtworks2021} are a revealing indicator for such technology movements. 

Developers in their role as applicants represent one side of RDD. On the other side, we find companies represented by their human resources (HR) departments.
While developers are focused on their own profiles, the company aims to create the best possible product using the right tools. These diverging interests can easily lead to conflicts in the choice of technologies for a specific project. The expertise they require may often involve technologies that developers are \textit{not} particularly keen to work with. Companies find themselves in a quandary here. 

On the side of companies and hiring professionals, RDD therefore implies that they benefit from using and specifically advertising popular technologies in their job postings that are more popular among developers.
That way, a larger number of applicants can be addressed, which may yield a better suited candidate in the end. Applicants, however, have no reason to assume that company job advertisements in reality do \textit{not} reflect the precise technological needs, and hence feel encouraged to optimize their profile even more with a wider range of current technologies.

It is very likely that such behaviors will eventually have negative side effects. First, it can affect applicants whose expectations are not met. 
Second, the long-term consequences for companies can be even more serious if technological heterogeneity and poor maintainability of created software result in high costs.
The legitimate question arises if there is more to RDD than anecdotal evidence, and whether these outlined interconnections actually exist in industrial practice.
We explored this question in a 2021 published study, which is briefly summarized in the following. Afterwards, we discuss potential consequences and provide practical recommendations for both perspectives.

\section{Does RDD really exist? The Need for Evidence}
\medskip

\noindent
Due to missing scientific research on this subject, we lacked a clear definition for the term \textit{Résumé-Driven Development}.

It can be sporadically found in blogs and discussion forums, where it often leads to controversial debates or even polemics~\cite{Loukides2014,HackerNews2017}. 
Our in 2020 initiated wide-ranging online survey invited both software and IT professionals as well as hiring professionals to share their experiences with us. Developers in their various roles, including students from computer science degree programs, have been surveyed as the \textit{applicant} group. The \textit{hiring} group consisted of employees from HR departments, team managers, and specialized headhunters in the IT industry.
The survey aimed to explain the phenomenon with scientific methods and determine its influencing factors.

With the support of a prominent German IT magazine, we collected a total of 591 responses, with the majority being located in Germany. Divided into the two perspectives of \textit{hiring} and \textit{applicant}, the answers resulted in a ratio of 130 to 558 (answering to both perspectives was possible). Demographic data showed a realistic distribution in terms of professional experience and company size. 
The dominant job role was software engineer, followed by manager, student, and architect. 

\section{A Survey: Hiring and Applicants}
\medskip

\begin{figure}[ht!]
    %  \centering
     \hspace{-0.2cm}
     \includegraphics[keepaspectratio=true, width=0.5\textwidth]{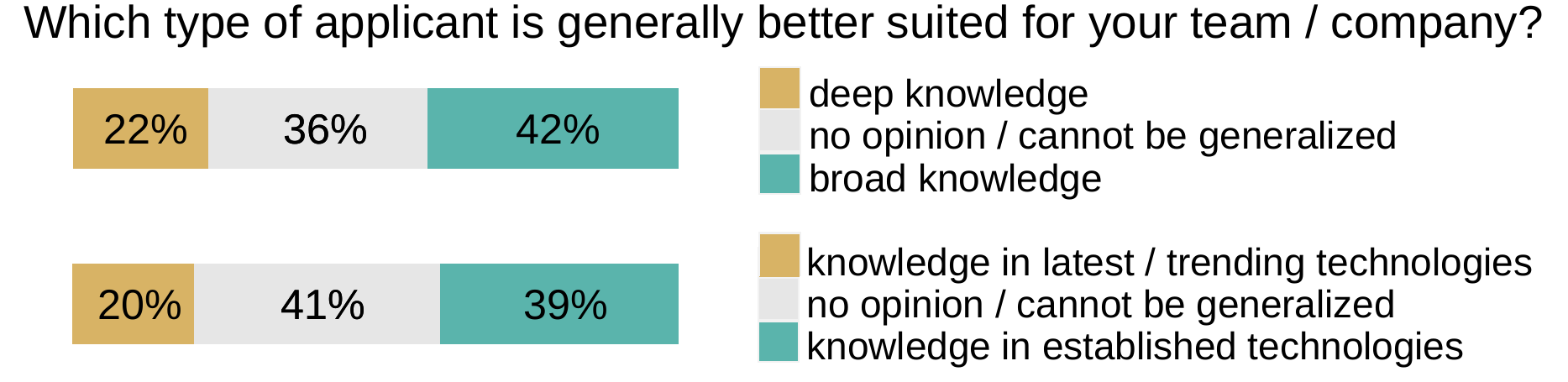}
     \caption{Preferred skills by Hiring professionals \cite{Fritzsch2021}}
     \label{fig:chart-hiring-prefer}
 \end{figure}

\noindent
The 130 \textbf{hiring participants} were asked regarding their preferences on knowledge and skills of \textit{applicants}, specifically the technology-related orientation \textit{breadth} vs. \textit{depth}. Around two thirds of the respondents considered both characteristics important. However, given a choice, 22\% of the participants would prefer a candidate with in-depth knowledge in specific technologies while to 42\% the candidate with broad knowledge in a variety of technologies appears more attractive (see Fig.~\ref{fig:chart-hiring-prefer}). This revealed a slight tendency towards a generalist profile in candidates. %WORDLIMIT candidates with broad knowledge in a variety of technologies appeared more attractive.

 \begin{figure*}[ht!]
    \centering
    \includegraphics[width=1.01\textwidth]{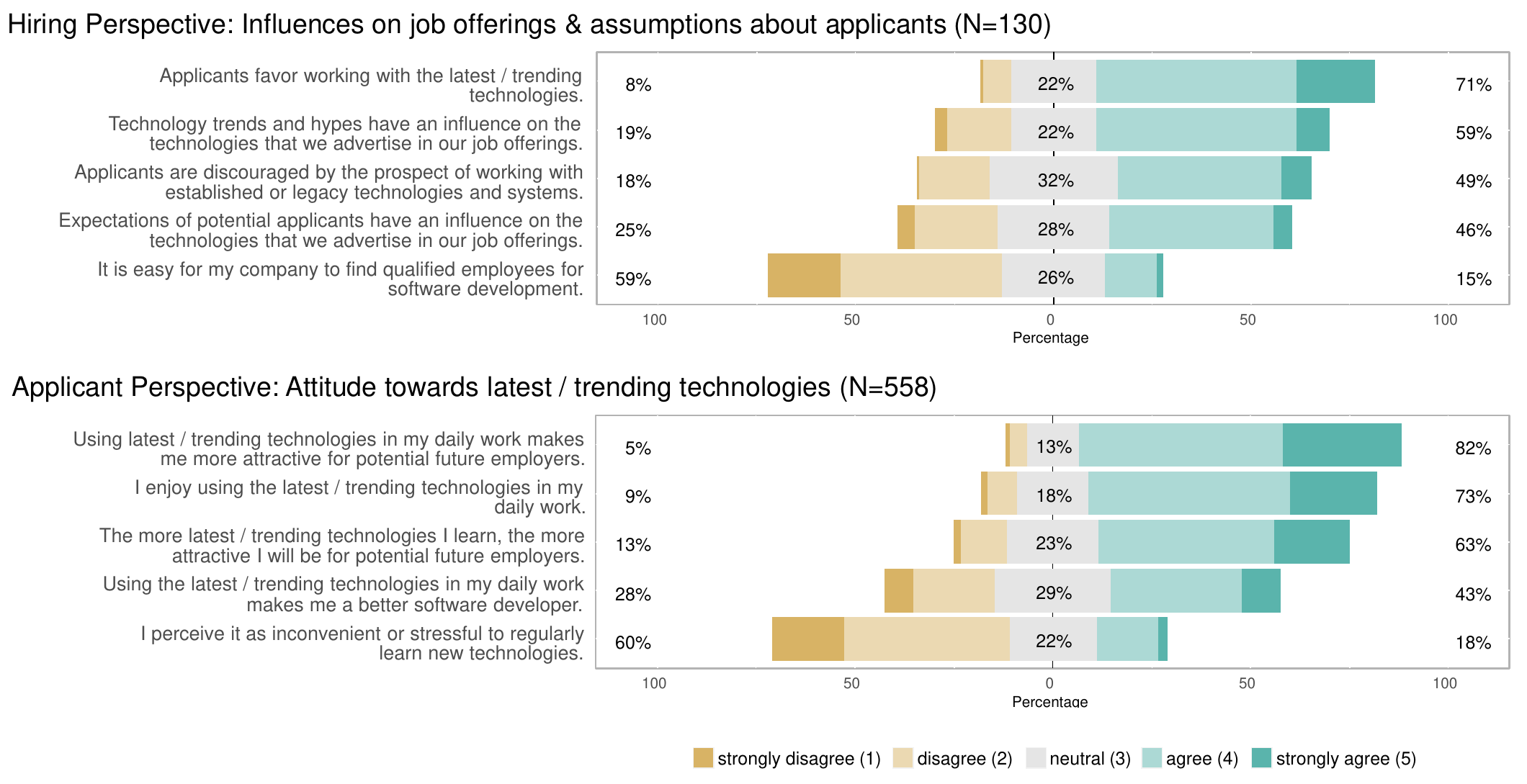}
    \caption{Survey results for Hiring and Applicant perspectives (condensed) \cite{Fritzsch2021}}
    \label{fig:chart-likert-results}
\end{figure*}

\noindent
Queried after the technology-related characteristics \textit{established} and \textit{latest/trending},
a significant advantage for knowledge in established technologies (39\%) over knowledge in current technology trends (20\%) became apparent. 41\% could or did not want to make a general statement on that (see Fig.~\ref{fig:chart-hiring-prefer}).
We conclude that broad knowledge in established technologies tends to more likely meet the needs of hiring professionals and thus companies. 

An important aspect in the interaction of both groups constitute the \textit{expectations from} / \textit{assessment of} the other side (see Fig.~\ref{fig:chart-likert-results}). 71\% of the hiring participants agreed that software developers would generally enjoy working with the latest/trending technologies. In addition, about half of the \textit{hiring} respondents assumed that applicants would even be discouraged by the prospect of working with established (\enquote{legacy}) technologies. Interestingly, the majority (59\%) of the \textit{hiring} respondents also agreed that technology trends have an impact on the contents of their own job advertisements. When asked directly, a large fraction of 46\% admitted that their advertised technologies are influenced by the expectations of potential applicants.

The 558 \textbf{applicant participants} were first asked regarding the role that technology trends play for them in general and second while creating their profile or CV (see Fig.~\ref{fig:chart-likert-results}). A large majority of 73\% stated that they enjoy using latest and trending technologies in their daily work. This matches the perception of the \textit{hiring} side, with an almost equal percentage. 
In addition to this intrinsic motivation for using trending technologies, 82\% of the \textit{applicants} were convinced that such knowledge and skills would make them more attractive for potential employers. Moreover, 63\% confirmed that an even higher variety of technologies would further increase their own attractiveness. However, only 42\% of the respondents believed that they would become better software developers by using such technologies.
The \textit{applicant} participants furthermore reported predominantly positive experiences with using trending technologies. 
Finally, one fifth of the respondents admitted that they have already used hyped technologies, although it was not the most appropriate choice for the concrete project or application.

These results suggest the existence of the RDD phenomenon on both sides. Technology trends do not always prove beneficial in practice, but are considered significantly more important when attracting applicants (\textit{hiring} perspective) and likewise increasing one's own attractiveness (\textit{applicant} perspective) in the hiring process.
The exploratory study culminated in the development of a theoretical construct that covers characteristics of both interacting groups, as well as strengthening predictors for its existence (see Fig. \ref{fig:chart-rdd-theory}). 
The \textit{hiring} perspective is characterized by two facets: the degree to which 1) technology trends and 2) the expectations of applicants influence their job offerings. Whereas, the presence of the \textit{applicant} perspective results from 1) the degree to which applicants are convinced that knowledge of trending technologies makes them more attractive for companies and 2) the importance of trends / hypes in the choice of technology. The thereby shaped RDD construct is reinforced by the strengthening predictors on the right side of the illustration in Fig. \ref{fig:chart-rdd-theory}.

Based on these findings, our empirical study proposes the following definition of RDD~\cite{Fritzsch2021}:

\begin{figure*}[htbp]
    \centering
    %\captionsetup{width=.95\linewidth}
    \includegraphics[width=1.0\textwidth]{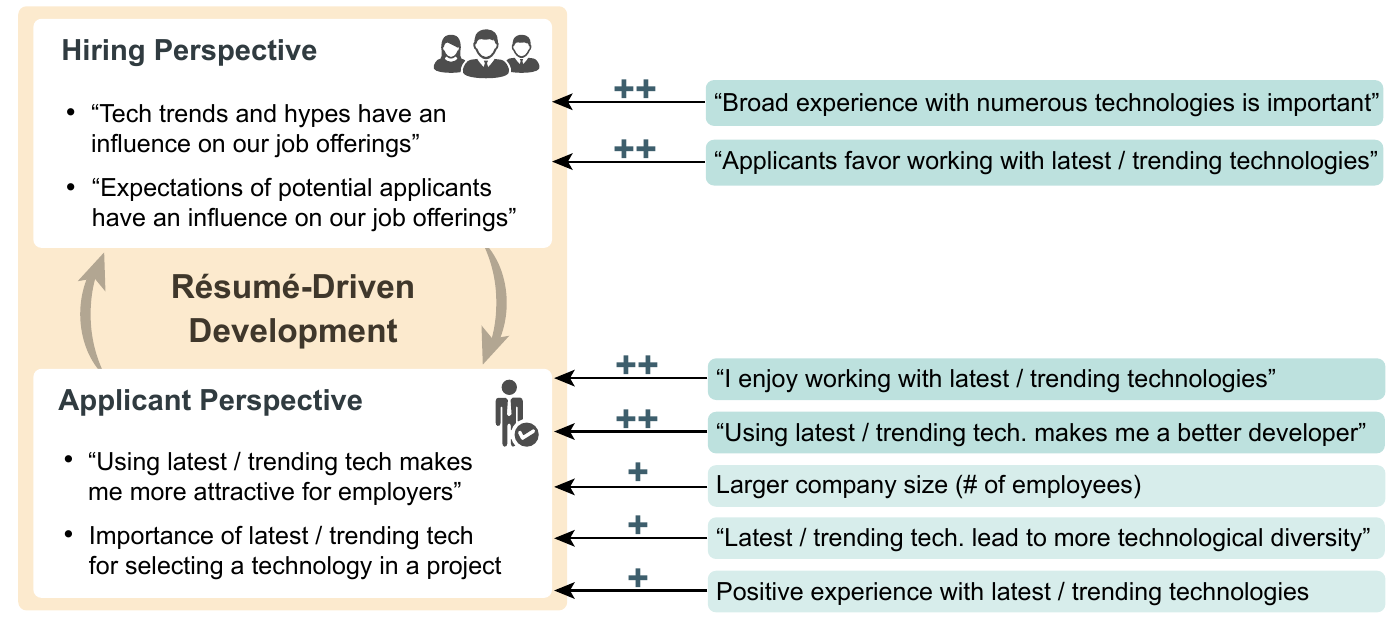}
    \caption{Theory of Résumé-Driven Development. HR and software professionals take the perspectives of \textit{hiring} and \textit{applicant} respectively, which are characterized by their facets in RDD. Strengthening predictors are linked to each perspective via + (moderate) and ++ (strong)~\cite{Fritzsch2021}. 
    %Icons from Noun Project\protect\footnotemark (Wilson Joseph; Creative Stall, PK).
   }
    \label{fig:chart-rdd-theory}
\end{figure*}

\vspace{0.1cm}
\begin{boxC}
\large
\setstretch{1.15}
\noindent
\enquote{\textit{Résumé-Driven Development (RDD) is an interaction between human resource and software professionals in the software development recruiting process. It is characterized by overemphasizing numerous trending or hyped technologies in both job advertisements and CVs, although experience with these technologies is actually perceived as less valuable on both sides. RDD has the potential to develop a self-sustaining dynamic.}}
\end{boxC}
\vspace{0.1cm}

\noindent
The definition and construct of \textit{Résumé-Driven Development} reflect this dynamic, based on a substantial part of the sample in the interview study. 
Further analysis found a positive correlation with the company size. However, the data did not show any significant correlation with the labor market situation, as collected from \textit{hiring} participants via their perceived difficulty in finding appropriate candidates.

\section{Consequences for Software Development Practice}
\medskip

\noindent
The question arises what potential effects \textit{Résumé-Driven Development} has on the practice of software development. Here, we can distinguish two main areas. 

First, long-term effects on software quality are very likely. The majority of participants on the \textit{applicant} side agreed that constantly emerging technology trends increase the diversity of languages, frameworks, and tools used in their company. This boosts the complexity~\cite{Ebert2017} and hence affects the maintainability of the developed software, either through the management of dependencies and updates or the necessary knowledge transfer within the team. In addition, a lack of reliability is often attributed to immature technologies. 
The consequences may not be immediately apparent, but manifest in the mid- to long-term. A recent article in IEEE Software on the subject of \textit{\enquote{The future of software development}} points to already visible consequences of \textit{\enquote{overwhelming complexity, combined with insufficient development competences}}, which lead to poor software quality and therefore, e.g., to the \textit{\enquote{public's decreasing acceptance of [\ldots] self-driving cars}}~\cite{Ebert2017}.

The second area of impact concerns the recruiting process itself. Here, RDD can arouse false expectations on both sides. Applicants generally dislike it if their future role in the company is not clearly defined. 
Inadequately communicated hiring criteria were identified as one of the main deficiencies in the analysis of over 10,000 job applicants' reviews on the Glassdoor career portal~\cite{Behroozi2020}. A subsequent high turnover is costly for both sides, but especially for the company: the cost-intensive training was a false investment and, in the worst case, the newcomer leaves behind code in a cutting-edge technology that none of the other team members can maintain. Existing studies directly connect a high turnover with increased knowledge loss~\cite{Nassif2017}, and reduced development productivity and software quality in general \cite{Bass2018}. 

Another point that was revealed by our study is the associated neglect of soft skills. These include social skills such as communication, self-motivation, and the ability to learn, as well as a basic understanding of the principles behind the various technologies.

\section{Recommendations for Companies}
\medskip

\noindent
The consequences of RDD can manifest in various ways.
To avoid ending up with the above outlined threats and hence risking severe long-term damage, we encourage hiring professionals and companies to consider the following points:

\begin{itemize}
	\item \textbf{Restrict the diversity of languages, frameworks, and tools used to an adequate and manageable extent.} Introducing a new technology should be well motivated and agreed on by lead developers or architects. A technology-related decision should always be justified by a concrete business need. Letting every developer select the tool of their choice for a given task is rarely the most sustainable approach.
	\item \textbf{The used technologies should always be familiar to several team members.} Just like for data and servers, it is highly recommended to have a backup for expertise too. It can happen unexpectedly fast that people with important knowledge are unavailable or change companies, leaving behind an expensive legacy. The \textit{truck factor}~\cite{Avelino2016}, i.e., the smallest number of people who need to get hit by a truck for the team to descend into knowledge and maintenance problems, describes this risk and suggests ways of mitigation.
	Well-documented decisions, tools, and processes can also help to prevent such knowledge loss. 
	\item \textbf{Communicate your needs and expectations early and clearly.} Enrich job advertisements only with technologies that are essential for the given position, and mark \textit{nice-to-have} requirements as such. This avoids frustration for the team and applicant that could otherwise lead to a quick turnover. A recent study on the technical interview process identifies such inadequately communicated criteria as a main deficiency expressed by applicants~\cite{Behroozi2020}.
	\item \textbf{Probe applicants critically after their motivations and intentions.} In the hiring process, be careful when assessing applicant profiles. Résumé-driven developers may not aim to stay in a company for long, but rather watch out for new opportunities, allowing them to enrich their profile.
	\item \textbf{Put equal emphasis on applicants' soft skills, and do not overemphasize technology-related knowledge}. While the latter can be beneficial to get up to speed quickly, it is eventually outweighed by more fundamental soft skills, e.g., when coping with difficult tasks or when collaborating in a cross-functional team. Hire for the long term.
	\item \textbf{Organize company-internal hackathons or coding challenges to allow developers to acquire and demonstrate new skills}~\cite{Pe-Than2019}. It offers room for experimenting with new technologies and thereby increases motivation. It can also be an opportunity to try out new approaches in a \enquote{sandbox} manner before risking false investments.
\end{itemize}

\section{Recommendations for Applicants}
\medskip

\noindent
Our survey confirmed that the vast majority of software professionals enjoy working with the latest / trending technologies. Indeed, it is a great way to keep yourself up-to-date, broaden your technical domain, and obtain a well-paid, fulfilling job. However, as you learn, do not get trapped in the RDD spiral. 
Consider the following food for thought to resist the mindset that primarily the latest technology on your resume is making you attractive to potential employers.

\begin{itemize}
    \item \textbf{You can hardly be an expert in all technologies, and this is perfectly fine.} Having a rich repertoire of skills to offer to employers is a great asset. But if we are honest, few of us are experts in all the technologies that we list in our resume. We all have personal preferences and expertise --- may it be that you are a Python guru or C++ enthusiast, you may probably not be a Java expert on top. Therefore, clearly state your skill level and experience with every listed technology to avoid misunderstandings in the hiring process. Our findings further suggest that there is no need to emphasize a variety of trending technologies, since companies are predominantly in demand for skills in established technologies. Promote these skills adequately.
    \item \textbf{Structure your resume by projects and professional experience rather than technology skills.} By adding the used technologies for each project and phase, you may demonstrate that you can easily learn new technologies in a professional context. This way, i.e., when seeing used technologies in context, hiring professionals can assess and compare your skill level more easily.
    \item \textbf{Read job advertisements with a little (healthy) skepticism.} Not all required skills for a particular position are really required. They are often comprehensive to address a large number of applicants. Figure out the main focus of this position and which other skills may have been merely added for decoration purposes. The latter can indeed be such latest / trending technologies that are actually not really needed. 
    \item \textbf{Pay attention to and strengthen your soft skills, just as you do for technologies you aim to master.} Software development is becoming less and less a discipline that builds on  individualists. Agile development practices are characterized by a high level of interaction that pushes social skills to the foreground. In the long term, these are at least equally important and even essential to pave the way when aiming to change your technical focus towards a management career path.
    \item \textbf{Create high-quality products by choosing the \emph{right} technologies.} Companies pay considerable salaries for good employees, but we need to understand that this comes from creating value for their customers. Customers expect products of high quality and efficiently working organizations when signing long-term contracts. Hence, the adequacy of chosen tools and methods should always be checked first in terms of this aspect. Keep in mind that it can be equally fulfilling (and less frustrating) for every developer on the team if the product is of high quality and thus easy to maintain.
    \item \textbf{After all, keep learning and improving.} There are great ways to stay ahead of the latest technological developments, even if they are no adequate choices for your current work project. If you are eager to learn technologies that are not currently used in your organization, there are often other ways to learn them than imposing them on your current project. Open-source projects, e.g., offer a great way to indulge in trying a new framework or language. It is also a great way to socialize and make new contacts with like-minded people. For those who prefer to develop their professional skills during working hours, encourage your company to think about a hackathon or offering further training opportunities.
\end{itemize}

\section{Conclusion}
\medskip

\noindent
There are always two sides to a coin. 
Hence, it is possible to derive both negative and positive aspects from the RDD phenomenon. 
The renowned computer scientist Donald Knuth may speak rather in favor of giving software developers the freedom to choose their preferred tools and technologies: \textit{\enquote{computer programming is an art […] A programmer who subconsciously views himself as an artist will enjoy what he does and will do it better.}}~\cite{Knuth1974}

The history of software development has always been characterized by permanent change, and the ever faster expanding technology landscape requires a constantly high level of willingness to learn. 
To both \textit{hiring} and \textit{applicants}, it is important to understand that such a thirst for knowledge is not demonstrated by a long list of trending technologies in one's resume. There are better ways of demonstrating profound knowledge and interest in the latest developments. Putting more emphasis on accomplished projects and phases enriched with the used technologies conveys a more professional approach.
Companies need to understand that this dynamic has potentially negative consequences for them as well. 
Clearly communicated needs and expectations in job advertisements that demand only those technologies being essential for the given position are an easy step towards avoiding false expectations and frustration later on. Appropriately screening applicant profiles also regarding their soft skills can help to avoid an overemphasis of technology-related knowledge.

We derive from the analyzed data that there is more than anecdotal evidence to the phenomenon of \textit{Résumé-Driven Development}, and believe that RDD has the potential for severe negative consequences.
Future research might draw a more precise picture by showing, e.g., how different industries and domains are affected by RDD. Long-term studies could yield more precise interconnections and insights, and hence result in specific guidance for practitioners.
For more details about the described survey, we refer the interested reader to our paper \enquote{Résumé-Driven Development: A Definition and Empirical Characterization} presented at the International Conference on Software Engineering (ICSE 2021)~\cite{Fritzsch2021}.

% References
\bibliographystyle{IEEEtran}
\bibliography{references.bib}

\begin{wrapfigure}{l}{0.10\textwidth}
\includegraphics[width=0.12\textwidth]{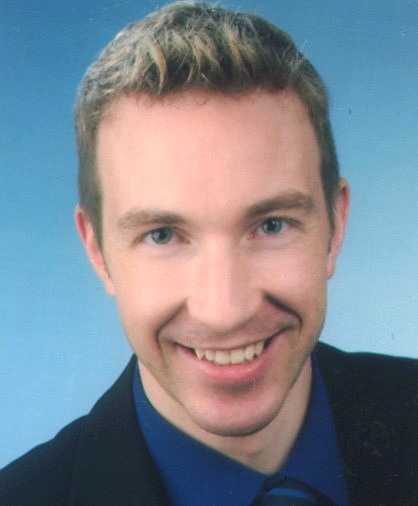}
\vspace{-30pt}
\end{wrapfigure}
\begin{IEEEbiography}{Jonas Fritzsch}{\,} is a researcher in the field of software engineering and architecture at the University of Stuttgart, Germany. He benefits from over fifteen years of experience in enterprise software development at his previous employer HPE (formerly HP). As a university lecturer, he teaches programming and algorithms in computer science courses. Contact him at \url{jonas.fritzsch@iste.uni-stuttgart.de}
\end{IEEEbiography}

\begin{wrapfigure}{l}{0.10\textwidth}
\vspace{-10pt}
\includegraphics[width=0.12\textwidth]{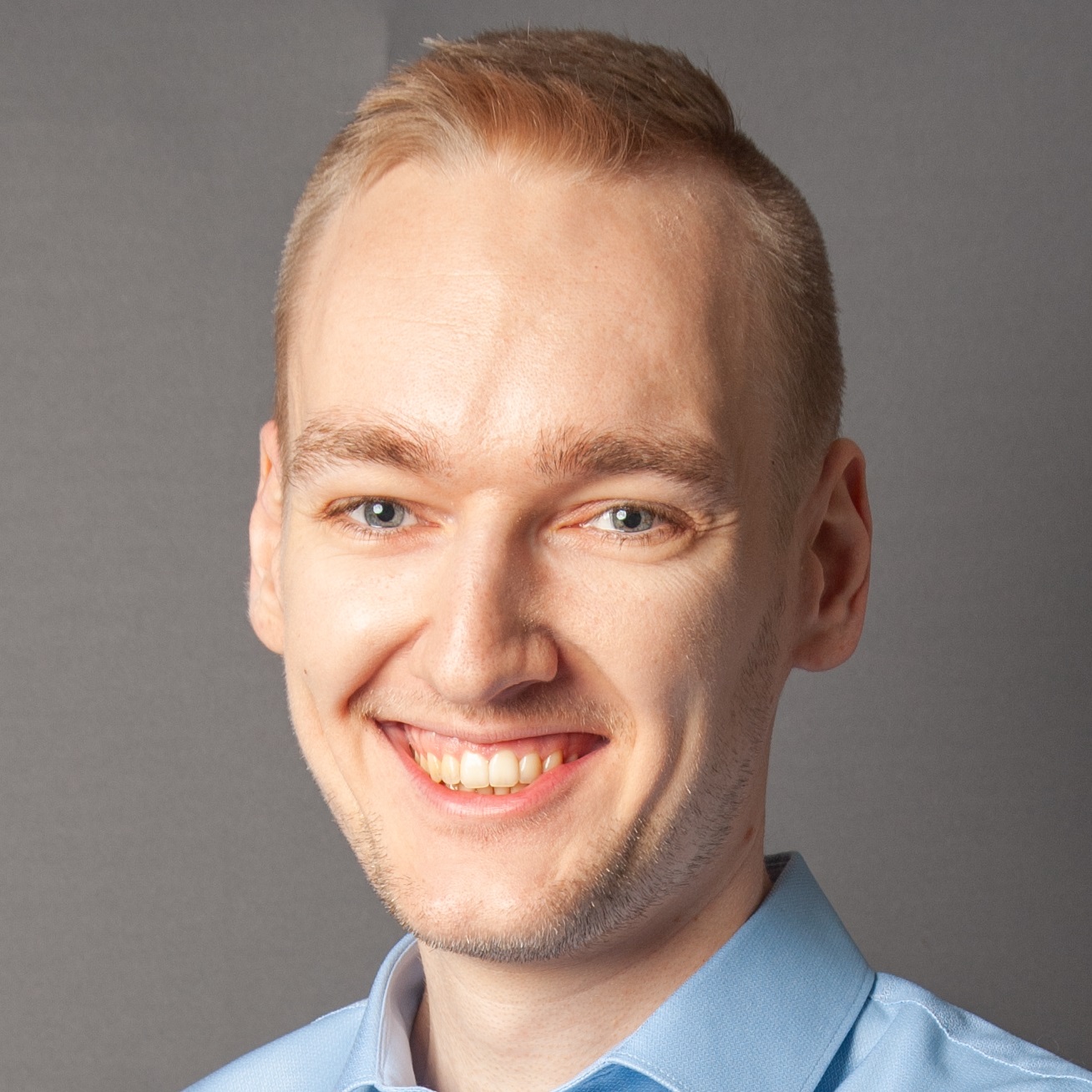}
\vspace{-30pt}
\end{wrapfigure}
\begin{IEEEbiography}{Marvin Wyrich}{\,}is a researcher at the University of Stuttgart, where he has been part of the empirical software engineering research group since 2018, and at Saarland University as part of the software engineering research group since 2023. His research interests include empirical and behavioral software engineering, with a focus on developing sound research methodologies. Contact him at \url{wyrich@cs.uni-saarland.de}
\end{IEEEbiography}

\begin{wrapfigure}{l}{0.10\textwidth}
\vspace{-10pt}
\includegraphics[width=0.12\textwidth]{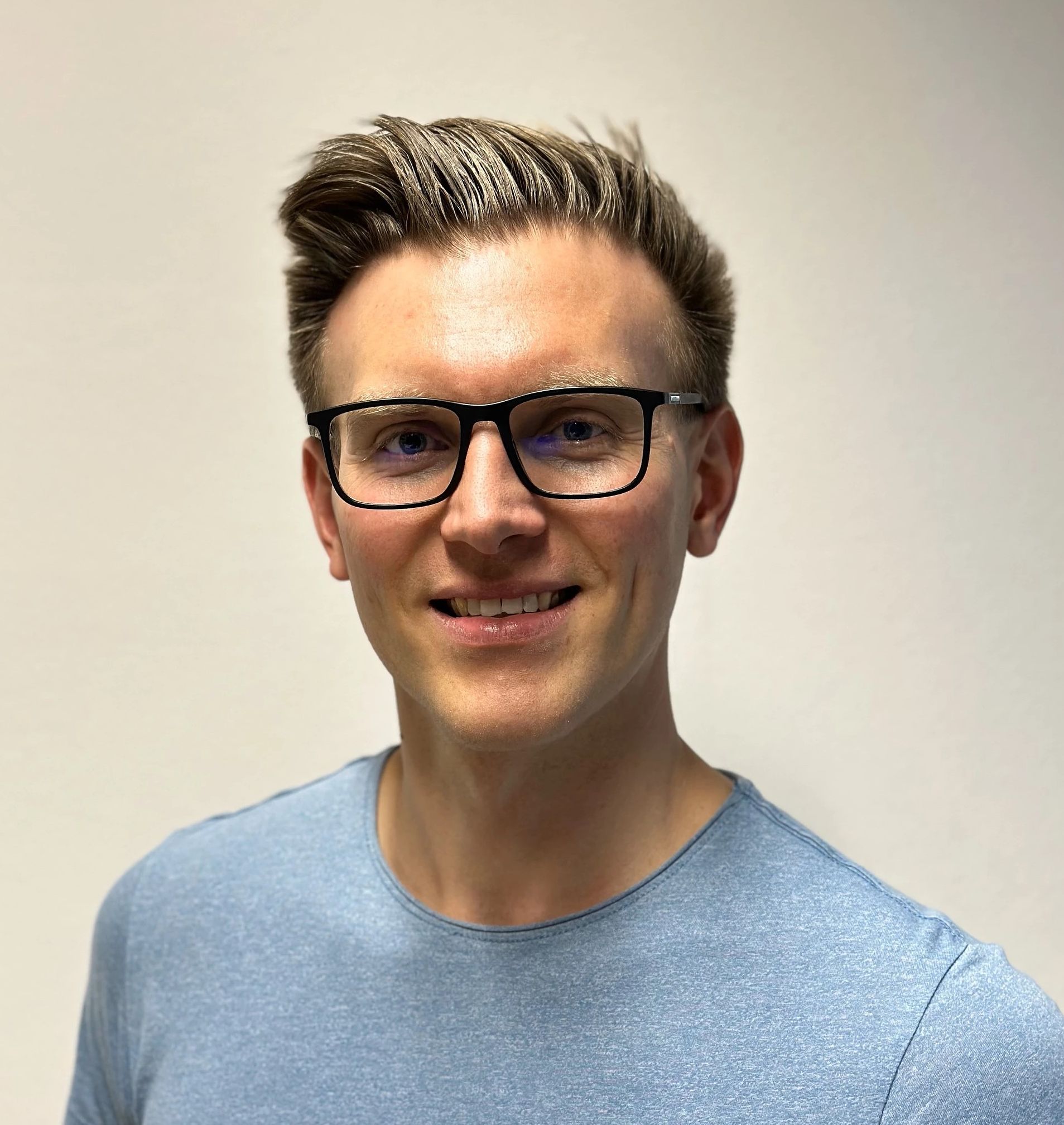}
\vspace{-30pt}
\end{wrapfigure}
\begin{IEEEbiography}{Justus Bogner}{\,} is a Postdoctoral Researcher at the University of Stuttgart, Germany.
Within the Empirical Software Engineering group, he leads the division \textit{Software Engineering for AI- and Microservice-based Systems}.
He has worked as a software engineer in industry for over 9 nears (HP, HPE, DXC Technology), building mostly Web- and service-based enterprise applications. Contact him at \url{justus.bogner@iste.uni-stuttgart.de}
\end{IEEEbiography}

\begin{wrapfigure}{l}{0.10\textwidth}
\vspace{-10pt}
\includegraphics[width=0.12\textwidth]{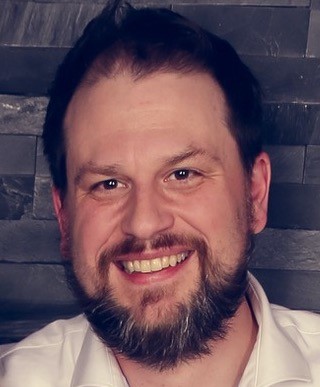}
\vspace{-30pt}
\end{wrapfigure}
\begin{IEEEbiography}{Stefan Wagner}{\,} is a Full Professor of empirical software engineering and director at the Institute of Software Engineering at the University of Stuttgart. His research interests are human aspects, software quality, automotive software, AI-based systems, and empirical studies. He studied computer science in Augsburg and Edinburgh and received a doctoral degree from the Technical University of Munich. He is a senior member of IEEE and ACM. Contact him at \url{stefan.wagner@iste.uni-stuttgart.de}.
\end{IEEEbiography}

\end{document}